\documentclass[12pt]{article}

\usepackage{amssymb}
\usepackage{amsmath}
\usepackage{amsfonts}

\newcommand{\C}{\mathbb{C}}
\newcommand{\Z}{\mathbb{Z}}

\newcommand{\be}{\begin{equation}}
\newcommand{\bea}{\begin{eqnarray}}
\newcommand{\eea}{\end{eqnarray}}
\newcommand{\nn}{\nonumber}
\newcommand{\kt}{\rangle}
\newcommand{\br}{\langle}
\newcommand{\cun}{\mbox{\footnotesize${\cal N}$}}

\newcommand{\ed}{\end{document}}
\newcommand{\bbr}{\br\!\br}
\newcommand{\kkt}{\kt\!\kt}

\begin{document}

\title{Pseudo-Supersymmetric Quantum Mechanics and Isospectral Pseudo-Hermitian Hamiltonians}
\author{Ali Mostafazadeh\thanks{E-mail address: amostafazadeh@ku.edu.tr}\\ \\
Department of Mathematics, Ko\c{c} University,\\
Rumelifeneri Yolu, 80910 Sariyer, Istanbul, Turkey}
\date{ }
\maketitle

\begin{abstract}
We examine the properties and consequences of pseudo-supersymmetry
for quantum systems admitting a pseudo-Hermitian Hamiltonian. We
explore the Witten index of pseudo-supersymmetry and show that
every pair of diagonalizable (not necessarily Hermitian)
Hamiltonians with discrete spectra and real or complex-conjugate
pairs of eigenvalues are isospectral and have identical degeneracy
structure except perhaps for the zero eigenvalue if and only if
they are pseudo-supersymmetric partners. This implies that
pseudo-supersymmetry is the basic framework for generating
non-Hermitian $PT$-symmetric and non-$PT$-symmetric Hamiltonians
with a real spectrum via a Darboux transformation, and shows that
every diagonalizable Hamiltonian $H$ with a discrete spectrum and
real or complex-conjugate pairs of eigenvalues may be factored as
$H=L^\sharp L$ where $L$ is a linear operator with pseudo-adjoint
$L^\sharp$. In particular, this factorization applies to
$PT$-symmetric and Hermitian Hamiltonians. The nondegenerate
two-level systems provide a class of Hamiltonians that are
pseudo-Hermitian. We demonstrate the implications of our general
results for this class in some detail.
\end{abstract}

\baselineskip=24pt

\section{Introduction}

In a recent series of papers \cite{pt1,pt2,pt3} we have reported a comprehensive study of the
necessary and sufficient conditions for the reality of the spectrum of a general (not necessarily Hermitian) diagonalizable Hamiltonian. We also elucidated the consequences of antilinear symmetries, such as $PT$-symmetry \cite{pt} -- \cite{kp}, on the spectral properties of a diagonalizable Hamiltonian. The essential ingredient of this analysis is the notion of a {\em pseudo-Hermitian operator}. In \cite{pt1} we also introduced a generalization of supersymmetric quantum mechanics \cite{susy,susy2} called {\em pseudo-supersymmetric quantum mechanics} that applied for $\Z_2$-graded quantum systems \cite{npb-01} with a pseudo-Hermitian Hamiltonian. The purpose of this article is to explore various properties and applications of pseudo-supersymmetry. In particular, we establish the fact that all the available methods of generating non-Hermitian Hamiltonians with a real spectrum by intertwining the latter with Hermitian Hamiltonians \cite{cannata-98,andrianov,znojil-2000,Bagchi-2000a,Bagchi-2000b,bmq,cannata-01,tkachuk-01,bagchi-02,milanovic-02} fit to the framework provided by pseudo-supersymmetry. We also point out the implications of pseudo-supersymmetry for ordinary unitary quantum mechanics.

In this article we shall only consider the Hamiltonians that are diagonalizable and have a discrete spectrum. By this we mean that $H$ admits a complete biorthonormal system of eigenvectors $\{|\psi_n,a\kt,|\phi_n,a\kt\}$. The latter satisfy the following defining properties \cite{biortho}.
     \bea
    &&H|\psi_n,a\kt=E_n|\psi_n,a\kt,~~~~H^\dagger|\phi_n,a\kt=E_n^*|\phi_n,a\kt\;,
    \label{o1}\\
    &&\br\phi_m,b|\psi_n,a\kt=\delta_{mn}\delta_{ab},
    \label{o2}\\
    && \sum_n\sum_{a=1}^{d_n}|\phi_n,a\kt\br\psi_n,a|
    =\sum_n\sum_{a=1}^{d_n}|\psi_n,a\kt\br\phi_n,a|=1,
    \label{o3}
    \eea
where a dagger stands for the adjoint of the corresponding
operator, $d_n$ is the multiplicity (degree of degeneracy) of the
eigenvalue $E_n$, $n$ is the spectral label, and $a$ and $b$ are
degeneracy labels.

The organization of the article is as follows. In Sections~2 and
~3, we review the basic properties of pseudo-Hermitian operators
and pseudo-supersymmetric quantum mechanics. In Section~4 we study
the Witten index of pseudo-supersymmetry. In Section~5, we show
that every diagonalizable pair of isospectral Hamiltonians are
related by an intertwining relation. We then specialize to the
case of pseudo-Hermitian isospectral pairs of Hamiltonians with
identical degeneracy structure except possibly for the zero
eigenvalue, and show that they are related by a
pseudo-supersymmetry transformation. In Section~6, we examine the
applications of the ideas developed in Section~5 for general
nondegenerate two-level Hamiltonians. Finally, in Section~7, we
summarize our main results and present our concluding remarks.

\section{Pseudo-Hermiticity and Its Consequences}

Let ${\cal H_\pm}$ be Hilbert spaces, $A:{\cal H}_+\to{\cal H}_-$
be a linear operator and $\eta_\pm:{\cal H}_\pm\to{\cal H}_\pm$ be
linear (or antilinear) Hermitian, invertible operators. Then the
{\em pseudo-adjoint} $A^\sharp:{\cal H}_-\to{\cal H}_+$ of $A$
with respect to $\eta_\pm$ is defined by \cite{pt1}
    \be
    A^\sharp:=\eta_+^{-1}A\;\eta_-.
    \label{p-adjoint}
    \end{equation}
Now suppose that ${\cal H}_+={\cal H}_-$ is the Hilbert space
${\cal H}$ of a quantum system with a Hamiltonian $H$ and
$\eta_-=\eta_+=:\eta$. Then $H$ is said to be {\em
pseudo-Hermitian with respect to $\eta$} if $H^\sharp=H$, i.e.,
    \be
    H^\dagger=\eta H\eta^{-1}.
    \label{p-hermiticity}
    \end{equation}
$H$ is said to be {\em pseudo-Hermitian} if there is a linear, Hermitian, invertible operator $\eta$
with respect to which $H$ is pseudo-Hermitian.

The basic properties of pseudo-Hermitian operators are given in \cite{pt1}. See also \cite{ahmed}.
The role of pseudo-Hermiticity in characterizing non-Hermitian Hamiltonians with a real spectrum
and the relation to antilinear symmetries are discussed in \cite{pt2} and \cite{pt3}.
In the following we summarize the main results of \cite{pt1,pt2,pt3}.
    \begin{itemize}
    \item[] {\bf Theorem~1}: The following are equivalent.
        \begin{enumerate}
        \item The eigenvalues of $H$ are either real or come in complex-conjugate pairs.
        \item $H$ is pseudo-Hermitian with respect to a linear Hermitian invertible operator
        $\eta:{\cal H}\to{\cal H}$.
        \item $H$ has an (antilinear) symmetry generated by an antilinear operator $\sigma$,
        $[H,\sigma]=0$.
        \end{enumerate}
    \item[] {\bf Theorem~2}: The following are equivalent.
        \begin{enumerate}
        \item $H$ has a real spectrum.
        \item $H$ is pseudo-Hermitian with respect to $O^\dagger O$ for some        linear  invertible operator $O:{\cal H}\to{\cal H}$.\footnote{There is a minor error in
        \cite{pt2}: With $O$ as defined in Equation~(9) of \cite{pt2}, $H$ is
        pseudo-Hermitian with respect to $(OO^\dagger)^{-1}$.}
        \item $H$ is related to a Hermitian operator $h:{\cal H}\to{\cal H}$ by a similarity
        transformation, namely
            \be
            H=O^{-1}h O,
            \label{similar}
            \end{equation}
        where $O:{\cal H}\to{\cal H}$ is a linear invertible operator.
        \end{enumerate}
    \end{itemize}

Given a pseudo-Hermitian Hamiltonian $H$ there are infinitely many $\eta$ satisfying
(\ref{p-hermiticity}). These can however be expressed in terms of a complete biorthonormal basis $\{|\psi_n,a\kt,|\phi_n,a\kt\}$ of $H$ and a set of Hermitian matrices $M^{(n)}$. To see this, we shall
use $n_0$ and $\pm\nu$ to denote the spectral label $n$ corresponding to eigenvalues with $0$ and $\pm$ imaginary part, respectively. Then in view of (\ref{o1}) -- (\ref{o3}) and the fact that the complex eigenvalues come in complex-conjugate pairs, i.e.,
$E_{-\nu}=E_\nu^*$, we have
    \be
    H=\sum_{n_0}\sum_{a=1}^{d_{n_0}}
        E_{n_0}|\psi_{n_0},a\kt\br\phi_{n_0},a|+
         \sum_{\nu} \sum_{a=1}^{d_{\nu}} \left(
        E_{\nu}|\psi_{\nu},a\kt\br\phi_{\nu},a|+
        E_{\nu}^*|\psi_{-\nu},a\kt\br\phi_{-\nu},a|\right).
    \label{H=}
    \end{equation}
In view of this relation, one can show that the most general linear, Hermitian, invertible operator $\eta$ satisfying (\ref{p-hermiticity}) has the {\em canonical form}
    \[  \eta=\sum_{n_0}\sum_{a,b=1}^{d_{n_0}}
        M^{(n_0)}_{ab}|\phi_{n_0},a\kt\br\phi_{n_0},b|+
            \sum_{\nu} \sum_{a,b=1}^{d_{\nu}}\left(
        M^{(\nu)}_{ab}|\phi_{\nu},a\kt\br\phi_{-\nu},b|+
        {M^{(\nu)}}^\dagger_{ab}|\phi_{-\nu},a\kt\br\phi_{\nu},b|\right),   \]
where $M^{(n_0)}_{ab}=\br\psi_{n_0},a|\eta|\psi_{n_0},b\kt$ and
$M^{(\nu)}_{ab}=\br\psi_{\nu},a|\eta|\psi_{-\nu},b\kt$ are respectively the entries of invertible $d_{n_0}\times d_{n_0}$ and $d_\nu\times d_\nu$ matrices $M^{(n_0)}$ and $M^{(\nu)}$. Clearly $M^{(n_0)}$ are Hermitian.

It is not difficult to check that under a change of basis: $|\psi_n,a\kt\to\sum_{b=1}^{d_n}V^{(n)}_{ba}|\psi_n,b\kt$, with $V^{(n)}_{ab}$ being the entries
of an arbitrary invertible matrix $V^{(n)}$ and $n=n_0,\nu\pm$, the matrices $M^{(n)}$ change according to
    \[M^{(n_0)}\to {V^{(n_0)}}^\dagger M^{(n_0)}{V^{(n_0)}},~~~~~~
    M^{(\nu)}\to {V^{(\nu)}}^\dagger M^{(\nu)}{V^{(-\nu)}}.\]
This in turn implies that one can make a change of basis, i.e., choose the matrices $V^{(n)}_{ab}$
so that after the transformation, $M^{(\nu)}$ are identity matrices and $M^{(n_0)}$ are
diagonal matrices with $+1$ or $-1$ diagonal entries.\footnote{The fact that one cannot remove
the minus signs form the diagonal entries of $M^{(n_0)}$ has been overlooked in \cite{pt1}. This does not however affect any of its conclusions, as none of the results of \cite{pt1,pt2,pt3} uses the erroneous assumption that one may transform $M^{(n_0)}$ to identity matrices. This is because
one can always choose to work with an $\eta$ for which $M^{(n_0)}$ are positive-definite matrices. In this case there is a basis transformation that maps them to identity matrices.}
The latter is based on the same argument as the one used in the derivation of the canonical form of
a nondegenerate real symmetric bilinear (or more generally a Hermitian sesquilinear) form in linear algebra \cite{gelfand}.

In view of the transformation properties of the matrices $M^{(n)}$, there is a biorthonormal basis of
eigenvectors of $H$ in which $\eta$ takes the form
    \be
    \eta=\sum_{n_0}\sum_{a=1}^{d_{n_0}}
        \sigma^{n_0}_{a}|\phi_{n_0},a\kt\br\phi_{n_0},a|+
            \sum_{\nu} \sum_{a=1}^{d_{\nu}}\left(
        |\phi_{\nu},a\kt\br\phi_{-\nu},a|+
        |\phi_{-\nu},a\kt\br\phi_{\nu},a|\right),
    \label{eta=}
    \end{equation}
where $\sigma^{n_0}_{a}\in\{-1,1\}.$ In other words we have the following.
    \begin{itemize}
    \item[] {\bf Proposition:} Let $H$ be a diagonalizable pseudo-Hermitian Hamiltonian with a discrete spectrum. Then up to the choice of eigenvectors of $H$, the linear, Hermitian, invertible operators with respect to which $H$ is pseudo-Hermitian are determined by a set of signs
$\sigma^{(n_0)}_a$ where $n_0$ labels the real eigenvalues of $H$, $a\in\{1,2,\cdots,d_{n_0}\}$, and $d_{n_0}$ is the multiplicity of the eigenvalue labeled by $n_0$.
    \end{itemize}

We wish to conclude this section with the following remark.

In \cite{pt3}, we show that every diagonalizable Hamiltonian $H$ with a discrete
spectrum is pseudo-Hermitian with respect to an antilinear Hermitian, invertible operator
${\cal T}$, i.e., $H$ is anti-pseudo-Hermitian. This result is essential for the proof of
Theorem~1, \cite{pt3}. In \cite{p46} we discuss a factorization property of symmetric matrices
that may be used to show that unlike $\eta$,  ${\cal T}$ is unique up to basis transformations.

\section{Pseudo-Supersymmetry}

Consider a Hilbert space ${\cal H}$ which is endowed with a pair of linear operators,
$\tau,{\cal Q}:{\cal H}\to{\cal H}$, satisfying
    \be
    \tau=\tau^\dagger=\tau^{-1},~~~~\{\tau,{\cal Q}\}=0.
    \label{t-q}
    \end{equation}
These are sufficient to conclude that $\tau$ is a $\Z_2$-grading operator, ${\cal H}$ is $\Z_2$-graded
\cite{npb-01}, i.e., ${\cal H}={\cal H}_+\oplus {\cal H}_-$ with
    \[{\cal H}_\pm:=\{|\psi\kt_\pm \in{\cal H}~|~\tau|\psi\kt_\pm=\pm|\psi\kt_\pm\},\]
and ${\cal Q}$ is an {\em odd operator}, i.e., ${\cal Q}$ maps ${\cal H}\pm$ to ${\cal H}\mp$.
Next, let $\eta:{\cal H}\to{\cal H}$ be a linear Hermitian invertible operator that commutes with $\tau$, i.e., $\eta$ is an {\em even operator}, and
    \be
    {\cal Q}^2={\cal Q}^{\sharp 2}=0,~~~~~~\{ {\cal Q},{\cal Q}^\sharp\}=2 H,
    \label{psusy}
    \end{equation}
where $\sharp$ is defined with respect to $\eta$. Equations~(\ref{psusy}), which in particular imply
$[{\cal Q},H]=0$, define the algebra of $N=2$ {\em pseudo-supersymmetric quantum mechanics}.\footnote{We use $N$ to denote the number of pseudo-Hermitian generators of pseudo-supersymmetry.} According to the last equation in (\ref{psusy}) the Hamiltonian $H$ is an even pseudo-Hermitian operator, and the quantum system has a {\em pseudo-supersymmetry} generated by ${\cal Q}$.

Following the ordinary supersymmetric quantum mechanics, we can obtain a simple two-component
realization of pseudo-supersymmetry in which the state vectors $|\psi\kt$, the grading operator $\tau$, the pseudo-supersymmetry generator ${\cal Q}$, the Hamiltonian $H$, and the operator $\eta$ are respectively represented as
    \bea
    |\psi\kt&=&\left(\begin{array}{c}
        |\psi\kt_+\\
        |\psi\kt_-\end{array}\right),~~~~
     \tau =\left(\begin{array}{cc}
       1& 0\\
        0 & -1 \end{array}\right),~~~~
    {\cal Q}=\left(\begin{array}{cc}
        0 & 0 \\
        D & 0 \end{array}\right),
      \label{p-susy-1}\\
    H&=&\left(\begin{array}{cc}
        H_+ & 0\\
        0 & H_- \end{array}\right),~~~~
    \eta=\left(\begin{array}{cc}
        \eta_+& 0\\
        0 & \eta_- \end{array}\right),
    \label{p-susy-2}
    \eea
where $|\psi\kt_\pm\in{\cal H}_\pm$, $D:{\cal H}_+\to{\cal H}_-$ is a linear operator,
$H_\pm:=H|_{{\cal H}_\pm}$, and $\eta_\pm:=\eta|_{{\cal H}_\pm}$. Note that in view of (\ref{psusy}) -- (\ref{p-susy-2}),
    \be
        H_+:=\frac{1}{2}\, D^\sharp D,~~~~
        H_-:= \frac{1}{2}\, D \,D^\sharp,
    \label{H-H}
    \end{equation}
where
    \be
        D^\sharp=\eta_+^{-1}D^\dagger\eta_-.
        \label{D=}
          \end{equation}
This follows from (\ref{p-susy-1}), (\ref{p-susy-2}), and
${\cal Q}^\sharp:=\eta^{-1}{\cal Q}^\dagger\eta$.

Moreover, it is not difficult to see that the pseudo-Hermiticity of $H$ with respect to $\eta$ implies
pseudo-Hermiticity of $H_\pm:{\cal H}_\pm\to {\cal H}_\pm$ with respect to $\eta_\pm$ and that
$H_\pm$ satisfy the intertwining relations
    \be
    D\,H_+=H_-D,~~~~D^\sharp H_-=H_+ D^\sharp\;.
    \label{inter}
    \end{equation}
As a consequence, $H_+$ and $H_-$ are isospectral, $D$ maps the eigenvectors of $H_+$ to those of $H_-$ and $D^\sharp$ does the converse, except for those eigenvectors that are eliminated by these operators. In \cite{pt1} we used this observation to construct a class of complex potentials with a real
spectrum. Other applications of the notion of pseudo-Hermiticity and pseudo-supersymmetry have been explored in \cite{kp}.

In view of the analogy with supersymmetry we shall call $H_+$ and $H_-$ {\em pseudo-superpartner Hamiltonians}. It is clear that as a result of the intertwining relations~(\ref{inter}) these Hamiltonians have the same multiplicity for their nonzero eigenvalues. Therefore, similarly to the case of superpartner Hamiltonians in ordinary supersymmetric quantum mechanics, $H_+$ and $H_-$ have identical degeneracy structure except possibly for the zero eigenvalue.

We end this section by the following two comments.
\begin{itemize}
\item[1.] One may easily generalize the algebra (\ref{psusy}) of $N=2$ pseudo-supersymmetry to the case where there are $\cun$ (non-pseudo-Hermitian) odd generators, ${\cal Q}_1,{\cal Q}_2,\cdots,{\cal Q}_{\cun}$. This yields the algebra of pseudo-supersymmetry of order $N=2\cun$, namely
    \be
    {\cal Q}_i^2={\cal Q}_i^{\sharp 2}=0,~~~~~
    \{ {\cal Q}_i,{\cal Q}_j^\sharp\}=2\delta_{ij} H,
    \label{cun}
    \end{equation}
where $i,j\in\{1,2,\cdots\cun\}$. Equivalently, one may introduce the $N$ pseudo-Hermitian generators:
    \[ Q^1_i:=\frac{{\cal Q}+{\cal Q}^\sharp}{\sqrt 2},~~~~~~
       Q^2_i:=\frac{{\cal Q}-{\cal Q}^\sharp}{\sqrt 2~i},\]
and express (\ref{cun}) in the form
    \[ \{ Q^\alpha_i,Q^\beta_j\}=\delta_{ij}\delta_{\alpha\beta} H,\]
where $\alpha,\beta\in\{1,2\}$ and $i,j\in\{1,2,\cdots\cun\}$.
\item[2.] Pseudo-supersymmetry of order $N$ has the same spectral degeneracy structure as the
supersymmetry of order $N$, i.e., the eigenspaces with nonzero eigenvalues are spanned by pairs
of pseudo-superpartner basis vectors $(|\psi\kt,{\cal Q}_i|\psi\kt)$. The only difference is that the spectrum of a pseudo-supersymmetric Hamiltonian $H$ may in general include negative as well as complex(-conjugate pairs) of eigenvalues.
\end{itemize}

\section{Witten Index of Pseudo-Supersymmetry}

Using the analogy with ordinary supersymmetry, we define the Witten index of pseudo-supersymmetry
according to
    \be
    \Delta:=d_0^{(+)}-d_0^{(-)},
    \label{witten}
    \end{equation}
where
    \bea
    d_0^{(+)}&:=&{\rm dim}~[{\rm ker}(H_+)]={\rm dim}~[{\rm ker}(D^\sharp D)],
    \label{d+}\\
    d_0^{(-)}&:=&{\rm dim}~[{\rm ker}(H_-)]={\rm dim}~[{\rm ker}(DD^\sharp)],
    \label{d-}
    \eea
and `dim' and `ker' abbreviate `dimension' and `kernel', respectively. As the eigenspaces of $H$ that are
associated with nonzero eigenvalues consist of pseudo-superpartner pairs of state vectors
$(|\psi\kt, D|\psi\kt)$, under continuous pseudo-supersymmetry preserving deformations of $H$ or
${\cal H}$ the Witten index is left invariant. Hence it is a topological invariant.

In order to reveal the mathematical meaning of the $\Delta$, we first observe that according to Equations (\ref{H-H}),
    \be
    {\rm ker}(D)\subseteq {\cal H}_+^{(0)}~~~~~{\rm and}~~~~~~
    {\rm ker}(D^\sharp)\subseteq {\cal H}_-^{(0)},
    \label{subset}
    \end{equation}
where
    \be
    {\cal H}_\pm^{(0)}:={\rm ker}(H_\pm)
    \label{zero-Hil}
    \end{equation}
is the eigenspace of $H_\pm$ with zero eigenvalue. It is not difficult to see that by virtue of (\ref{H-H}), $D$ maps ${\cal H}_+^{(0)}$ to ${\cal H}_-^{(0)}$ and $D^\sharp$ does the converse. This suggests the
definition of the restriction of $D$ and $D^\sharp$ to ${\cal H}_+^{(0)}$ and ${\cal H}_-^{(0)}$,
namely
    \be
    D_0:=D|_{{\cal H}_+^{(0)}}:{\cal H}_+^{(0)}\to {\cal H}_-^{(0)},~~~~~~~~~~~~
    D^\flat_0:=D^\sharp|_{{\cal H}_-^{(0)}}:{\cal H}_-^{(0)}\to {\cal H}_+^{(0)}.
    \label{D-D}
    \end{equation}
It is not difficult to see that in view of (\ref{subset}) and (\ref{D-D}),
    \bea
    {\rm ker}(D)&=&{\rm ker}(D_0),~~~~~~{\rm ker}(D^\sharp)={\rm ker}(D_0^\flat),
    \label{D-D-zero}\\
    D_0^\flat D_0&=&(D^\sharp D)|_{{\cal H}_+^{(0)}}=0,~~~~~~
    D_0D_0^\flat=(D D^\sharp)|_{{\cal H}_-^{(0)}}=0.
    \label{D-D-zero-1}
    \eea

It turns out that $D^\flat_0\neq D_0^\sharp$. In order to clarify the role of $D^\flat_0$, we
introduce\footnote{Here we also make use of the fact that $\tilde D_0^\dagger$ maps
$\tilde{\cal H}_-^{(0)}$ to $\tilde{\cal H}_+^{(0)}$.}
    \bea
    \tilde{\cal H}_\pm^{(0)}&:=&\eta_\pm{\cal H}_\pm^{(0)}:=
    \left\{\eta_\pm|\psi\kt\in{\cal H}_\pm~|~|\psi\kt\in{\cal H}_\pm^{(0)}\right\},
    \label{tilde-Hil}\\
    \tilde D_0^\dagger&:=&D^\dagger|_{\tilde{\cal H}_-^{(0)}}
            :\tilde{\cal H}_-^{(0)}\to\tilde{\cal H}_+^{(0)}.
    \label{tilde-D}
    \eea
Then, we can check that the diagram
    \[\begin{array}{ccc}
    {\cal H}_-^{(0)} & \stackrel{D_0^\flat}{\longrightarrow} & {\cal H}_+^{(0)}\\
    \eta_-\mbox{\Huge$\downarrow$} & \bigcirc & \mbox{\Huge$\uparrow$}\eta^{-1}_+\\
    \tilde{\cal H}_-^{(0)} & \stackrel{\tilde D_0^\dagger}{\longrightarrow}&
     \tilde{\cal H}_+^{(0)} \end{array}\]
is commutative, i.e.,
    \be
    D^\flat_0=\eta_+^{-1}\tilde D_0^\dagger\eta_-,
    \label{D-flat}
    \end{equation}
and that
    \be
    {\rm ker}(D^\dagger)={\rm ker}(\tilde D^\dagger_0)\subseteq\tilde{\cal H}_-^{(0)}.
    \label{ker-ker}
    \end{equation}

Next, consider the operators
    \be
    A_+:=\eta_- D_0:{\cal H}_+^{(0)}\to \tilde{\cal H}_-^{(0)},~~~~~~~~~~
    A_-:=\eta_+^{-1}\tilde D^\dagger_0:\tilde{\cal H}_-^{(0)}\to{\cal H}_+^{(0)}.
    \label{A}
    \end{equation}
Then in view of (\ref{H-H}), (\ref{D-flat}), (\ref{tilde-D}) and (\ref{zero-Hil}), we have $A_+A_-=0$
and $A_-A_+=0$. This implies that the sequence
    \be
    \Sigma:~~~~{\cal H}_+^{(0)}\stackrel{A_+}{\longrightarrow}
    \tilde{\cal H}_-^{(0)}\stackrel{A_-}{\longrightarrow} {\cal H}_+^{(0)}
    \stackrel{A_+}{\longrightarrow}\tilde{\cal H}_-^{(0)}
    \label{sigma}
    \end{equation}
is a complex. The Betti numbers $b_\pm$ and the analytic index of $\Sigma$ are respectively defined
by
    \bea
    &&b_\pm:={\rm dim}[{\rm ker}(A_\pm)/{\rm Im}(A_\mp)]=
    {\rm dim}[{\rm ker}(A_\pm)]-{\rm dim}[{\rm Im}(A_\mp)],
    \label{betti}\\
    &&{\rm Analytic~Index}(\Sigma):=b_+-b_-,
    \label{index}
    \eea
where `Im' abbreviates `Image'.

Now, in view of the fact that $\eta_\pm$ are invertible operators (vector-space isomorphisms) and
using Equations (\ref{D-flat}) and (\ref{A}), we have
    \bea
    &&{\rm dim}[{\rm ker}(A_+)]={\rm dim}[{\rm ker}(D_0)],~~~~~~~~~
    {\rm dim}[{\rm ker}(A_-)]={\rm dim}[{\rm ker}(D_0^\flat)],\nn\\
    &&{\rm dim}[{\rm Im}(A_+)]={\rm dim}[{\rm Im}(D_0)],~~~~~~~~~
    {\rm dim}[{\rm Im}(A_-)]={\rm dim}[{\rm Im}(D_0^\flat)].\nn
    \eea
Hence,
    \bea
    {\rm Analytic~Index}(\Sigma)&=&{\rm dim}[{\rm ker}(D_0)]+{\rm dim}[{\rm Im}(D_0)]
    -\left\{{\rm dim}[{\rm ker}(D_0^\flat)]+{\rm dim}[{\rm Im}(D_0^\flat)]\right\}\nn\\
    &=& d_0^{(+)}-d_0^{(-)}.
    \label{index=}
    \eea
Here in the last equation we have made use of the fact that $D_0$ and $D_0^\flat$ are linear
maps relating $d_0^{(\pm)}$-dimensional vector spaces ${\cal H}_\pm^{(0)}$, so that
    \bea
    {\rm dim}[{\rm ker}(D_0)]+{\rm dim}[{\rm Im}(D_0)]&=&d_0^{(+)},\nn\\
    {\rm dim}[{\rm ker}(D_0^\flat)]+{\rm dim}[{\rm Im}(D_0^\flat)]&=&d_0^{(-)}.\nn
    \eea
Equations~(\ref{witten}) and (\ref{index=}) show that the Witten index of pseudo-supersymmetry
coincides with the analytic index of a complex of Fredholm operators,
    \be
    \Delta={\rm Analytic~Index}(\Sigma).
    \label{witten=3}
    \end{equation}

Next, we consider the special case where the restriction of $\eta_\pm$ onto ${\cal H}_\pm^{(0)}$ is positive-definite (or negative-definite), i.e., for all nonzero $|\psi_\pm\kt\in{\cal H}_\pm^{(0)}-\{0\}$,
$|\bbr\psi_\pm|\psi_\pm\kkt_\pm|:=|\br\psi_\pm|\eta_\pm|\psi_\pm\kt|>0$.
This means that $H_\pm$ have no null eigenvectors with zero eigenvalue, or equivalently there is a complete biorthonormal basis of ${\cal H}_\pm$ in which $\eta_\pm$ has the canonical form~(\ref{eta=}) with arbitrary signs $\sigma_n^a$ for all nonzero eigenvalues $E_n$ and only positive (or negative) signs for the zero eigenvalue (if there is any). Now, consider an element
$|\psi_+\kt\in{\cal H}_+^{(0)}$ and let $|\psi_-\kt:=D_0|\psi_+\kt=D|\psi_+\kt$. Then, in view of
(\ref{zero-Hil}) and (\ref{D-flat}), we have
    \[ \bbr\psi_-|\psi_-\kkt_-=\br\psi_-|\eta_-|\psi_-\kt=
    \br\psi_+|D^\dagger\eta_-D|\psi_+\kt=
    \br\psi_+|\eta_+D^\sharp D|\psi_+\kt=0.\]
Because we consider the case where ${\cal H}_+^{(0)}$ has no nonzero null elements, this calculation
shows that $D_0|\psi_+\kt=|\psi_-\kt=0$, i.e., $|\psi_+\kt$ belongs to ker$(D_0)$. Therefore,
${\rm ker}(D_0)={\cal H}_+^{(0)}$ and $d_0^{(+)}={\rm dim}[{\rm ker}(D_0)]$. We can similarly show that if ${\cal H}_+^{(0)}$ includes no nonzero null vectors, then
${\rm ker}(D_0^\flat)={\cal H}_-^{(0)}$ and $d_0^{(+)}={\rm dim}[{\rm ker}(D_0^\flat)]$. Therefore,
in this case, the Witten index takes the form
    \be
    \Delta={\rm dim}[{\rm ker}(D_0)]-{\rm dim}[{\rm ker}(D_0^\flat)].
    \label{del}
    \end{equation}
Next, we employ Equations~(\ref{D-D-zero}) and recall that because $D^\sharp=
\eta_+^{-1}D^\dagger\eta_-$ and $\eta_\pm$ are invertible operators,
${\rm ker}(D^\sharp)={\rm ker}(D^\dagger)$. These together with (\ref{del}) lead to
    \be
    \Delta={\rm dim}[{\rm ker}(D)]-{\rm dim}[{\rm ker}(D^\sharp)]=
    {\rm dim}[{\rm ker}(D)]-{\rm dim}[{\rm ker}(D^\dagger)]=:{\rm Analytic~Index}(D).
    \label{witten=2}
    \end{equation}
Hence, in this case the Witten index is identical with the analytic index of $D$.

As we pointed out, the analysis leading to~(\ref{witten=2}) applies if $H_+$ and $H_-$ have no null eigenvectors with zero eigenvalue.\footnote{In this case, the method pursued in \cite{npb-02} fails
to apply.} However, as we show in Section~5 the Hamiltonians $H_+$ and $H_-$
always admit a factorization of the form
    \be
    H_+=L^\sharp L,~~~~~~~H_-=L\,L^\sharp,
    \label{L-L}
    \end{equation}
where $L:{\cal H}_+\to{\cal H}_-$ is a linear operator $L:{\cal H}_+\to{\cal H}_-$ whose
pseudo-adjoint $L^\sharp:=\eta_+^{'-1}L^\dagger\eta'_-$ is defined in terms of linear Hermitian invertible operators $\eta'_\pm$ with the property that in the pseudo-norm defined by $\eta_\pm'$ none of the eigenvectors of $H_\pm$ with zero eigenvalue is null. Therefore, the argument of the preceding paragraph holds and we can always express the Witten index as the analytic index of $L$,
    \be
    \Delta={\rm Analytic~Index}(L).
    \label{witten=20}
    \end{equation}

\section{Intertwining Isospectral Hamiltonians}

Consider two isospectral Hamiltonians $H_1$ and $H_2$ acting in Hilbert spaces
${\cal H}_1$ and ${\cal H}_2$ respectively. Let $\{|\psi_n^{(i)},a\kt, |\phi_n^{(i)},a\kt\}$ be a complete biorthonormal system associated with $H_i$ and $\Lambda^{(i)}_n:=\sum_{a_i=1}^{d^{(i)}_n}|\psi_n^{(i)},a_i\kt\br\phi_n^{(i)},a_i|$
where $d_n^{(i)}$ is the multiplicity of $E_n$ as an eigenvalue of $H_i$ and $i\in\{1,2\}$. Then according to (\ref{o1}) - (\ref{o3}), we have
    \be
    \sum_n\Lambda^{(i)}_n= 1~~~~{\rm and}~~~~
    H_i=\sum_n E_n \Lambda^{(i)}_n.
    \label{hi=}
    \end{equation}

Next let $\mu_n$ denote the smallest of $d_n^{(1)}$ and $d_n^{(2)}$ and introduce the operators $L_n:{\cal H}_1\to{\cal H}_2$ and $L(\alpha):{\cal H}_1\to{\cal H}_2$ according to
    \be
    L_n:=\sum_{a=1}^{\mu_n} |\psi_n^{(2)},a\kt\br\phi^{(1)}_n,a|
    ~~~~{\rm and}~~~~L(\alpha):=\sum_n \alpha_nL_n,
    \label{L}
    \end{equation}
where $\alpha=\{\alpha_n\}$ is an arbitrary (finite or infinite) sequence of complex numbers $\alpha_n$. Then one can easily show that
    \be
    L_m\Lambda^{(1)}_n=\delta_{mn}L_n=\Lambda^{(2)}_m L_n.
    \label{s}
    \end{equation}
Note that here there is no summation over repeated indices. Equations (\ref{s}) together with (\ref{hi=}) yield the intertwining relation:
    \be
    L(\alpha) H_1=H_2L(\alpha).
    \label{intertwine}
    \end{equation}
This is a demonstration of the fact that any two (diagonalizable) isospectral Hamiltonians are related by a Darboux transformation.

Next, suppose that the eigenvalues of one of the Hamiltonians (and consequently the other's) are real or come in complex-conjugate pairs and that $H_1$ and $H_2$ have the same spectral degeneracy structure.\footnote{This means that for all $n$, $d^{(1)}_n=d^{(2)}_n$.} Then according to Theorem~1, $H_1$ and $H_2$ are pseudo-Hermitian with respect to some linear Hermitian invertible operators $\eta_1$ and $\eta_2$, respectively,
    \be
    H_i^\dagger=\eta_i H\eta_i^{-1}.
    \label{p-her-hi}
    \end{equation}
In view of the results of Section~2, namely Equation~(\ref{eta=}), $\eta_i$ have the canonical form
    \be
    \eta_i=\sum_{n_0}\sum_{a=1}^{d_{n_0}}
        \sigma^{(i)n_0}_{a}|\phi^{(i)}_{n_0},a\kt\br\phi^{(i)}_{n_0},a|+
            \sum_{\nu} \sum_{a=1}^{d_{\nu}}\left(
        |\phi^{(i)}_{\nu},a\kt\br\phi^{(i)}_{-\nu},a|+
        |\phi^{(i)}_{-\nu-},a\kt\br\phi^{(i)}_{\nu},a|.\right),
    \label{eta_i=}
    \end{equation}
where $\{|\psi^{(i)}_n,a\kt,|\phi^{(i)}_n,a\kt\}$ is an appropriate complete biorthonormal eigenbasis of $H_i$ and $\sigma^{(i)n_0}_{a}\in\{-1,+1\}$.

Next, make the following choice for the signs $\sigma^{(i)n_0}_a$ and the complex numbers $\alpha_n$.
    \bea
    \sigma^{(1)n_0}_a&=&\left\{\begin{array}{ccc}
    -1&{\rm for}& E_{n_0}<0\\
    +1&{\rm otherwise,}&\end{array}\right.
    \label{s1}\\
    \sigma^{(2)n_0}_a&=&+1,
    \label{s2}\\
    \alpha_{n_0}&=&\sqrt{|E_{n_0}|},~~~~~~~
    \alpha_\nu=E_\nu,~~~~~~\alpha_{-\nu}=1.
    \label{alpha=}
    \eea
Then, a rather lengthy but straightforward calculation shows that for these choices of $\eta_i$ and
$\alpha=\{\alpha_n\}$, namely (\ref{eta_i=}) -- (\ref{alpha=}), we have
    \be
    H_1=L(\alpha)^\sharp L(\alpha)~~~~~{\rm and}~~~~~~
    H_2=L(\alpha)L(\alpha)^\sharp.
    \label{eq}
    \end{equation}
Now, consider a slightly more general case where the $H_1$ and $H_2$ have identical degeneracy structure except possibly for the zero eigenvalue.\footnote{Obviously this applies when zero belongs to their common spectrum.} In this case, we can easily check that Equations~(\ref{eq}) still hold. Therefore, in view of (\ref{eq}) and the results of Section~3, we have established the following.
    \begin{itemize}
    \item[] {\bf Theorem~3}: Let ${\cal H}_i$, with $i\in\{1,2\}$, be Hilbert spaces, and
$H_i:{\cal H}_i\to{\cal H}_i$ be diagonalizable linear operators with discrete spectra and real or complex-conjugate pairs of eigenvalues. Then ${\cal H}_1$ and ${\cal H}_2$ are isospectral and
have identical spectral degeneracy except perhaps for the zero eigenvalue if and only if
there is a linear operator $L:{\cal H}_1\to{\cal H}_2$ and linear, Hermitian, invertible operators $\eta_i:{\cal H}_i\to{\cal H}_i$ satisfying
    \be
    H_1=L^\sharp L~~~~~~~{\rm and}~~~~~~~~H_2=L L^\sharp,
    \label{thm3}
    \end{equation}
where $L^\sharp:=\eta_1^{-1}L^\dagger\eta_2$.
    \end{itemize}
Setting $D=\sqrt 2~L$ and comparing Equations (\ref{thm3}) and (\ref{H-H}), we see that $H_1$ and $H_2$ are pseudo-superpartner Hamiltonians. This is the main result of this article. It implies, in particular, that all the attempts in the literature \cite{cannata-98,andrianov,znojil-2000,Bagchi-2000a,Bagchi-2000b,bmq,cannata-01,tkachuk-01,bagchi-02,milanovic-02} to generate non-Hermitian Hamiltonians with a real spectrum by intertwining a Hermitian Hamiltonian fit to the framework provided by pseudo-supersymmetry.

The above result also applies to Hermitian Hamiltonians that have real eigenvalues. Specifically, we have the following corollary of Theorem~3.
    \begin{itemize}
    \item[]{\bf Corollary}: Every diagonalizable Hamiltonian $H$ that acts in a Hilbert space ${\cal H}$ and has a discrete spectrum and real or complex-conjugate pairs of eigenvalues (in particular every  Hermitian Hamiltonian with a discrete spectrum) may be factored as
    \be
    H=L^\sharp L
    \label{factor}
    \end{equation}
for some linear operator $L$ whose pseudo-adjoint $L^\sharp=\eta_1^{-1}L\eta_2$ is defined in terms
of a pair of linear, Hermitian, invertible operators $\eta_1,\eta_2:{\cal H}\to{\cal H}$ with respect to which $H$ is pseudo-Hermitian.
    \end{itemize}

\section{Application: Nondegenerate Two-Level Systems}

Two-level systems with Hermitian and non-Hermitian Hamiltonians have been extensively studied
for their various applications in modeling the interactions of spin 1/2 particles, the description of resonant states, and in particular in polarization optics and quantum computation. A large part of the literature on non-Hermitian Hamiltonians consist of the study of non-Hermitian two-level Hamiltonians.
See for example \cite{dtm-90,kp-91,mh,jmp-99} and references therein.

A nondegenerate two-level system has $\C^2$ as its Hilbert space and a linear map $H:\C^2\to\C^2$ as its Hamiltonian. Clearly in a basis $\{|1\kt,|2\kt\}$ of $\C^2$, which is usually assumed to be orthonormal, $H$ has the form of a $2\times 2$ matrix. It is well-known that one can perform a time-dependent (canonical) phase transformation~\cite{nova} that sets the trace of $H$ to zero. Therefore, without loss of generality one may suppose that $H$ has the form
    \be
    H=\left(\begin{array}{cc}
    a & b\\
    c & -a\end{array}\right),
    \label{2-H}
    \end{equation}
where $a,b,c\in\C$. The eigenvalues $E_n$ and a complete biorthonormal system
$\{|\psi_n\kt,|\phi_n\kt\}$ for $H$ are given by \cite{jmp-99}
    \bea
    &&E_1=-E,~~~~~~~E_2=E,
    \label{eg-va}\\
    &&|\psi_1\kt=\left(\begin{array}{c}
    -b\\
    a+E\end{array}\right),~~~~~~~
    |\psi_2\kt=\left(\begin{array}{c}
    a+E\\
    c\end{array}\right),
    \label{psi=}\\
    &&|\phi_1\kt=\frac{1}{N^*}\,\left(\begin{array}{c}
    -c^*\\
    a^*+E^*\end{array}\right),~~~~~~~
    |\phi_2\kt=\frac{1}{N^*}\,\left(\begin{array}{c}
    a^*+E^*\\
    b^*\end{array}\right),
    \label{phi=}
    \eea
where $E:=\sqrt{a^2+bc}$ has a nonnegative real part, $N:=2E(a+E)$, and we assume without loss of generality that $E\neq 0$, for otherwise either $H$ vanishes identically or it is not diagonalizable. Clearly the above choice of $\{|\psi_n\kt,|\phi_n\kt\}$ is valid provided that $E\neq -a$. If $E=-a$, we can always change the value of $a$ by performing a unitary transformation of the basis $\{|1\kt,|2\kt\}$ in terms of which $H$ has the form (\ref{2-H}). Because $E$ is an eigenvalue of $H$, this transformation leaves it invariant. In this way after  the transformation, $H$ has the same form
(\ref{2-H}) but the transformed $a$ satisfies $E\neq -a$. Equivalently, for $E=-a$ we can select a new set of complete biorthonormal system for $H$. In the following, we shall assume that $E\neq -a$.

One can easily check that (\ref{psi=}) and (\ref{phi=}) satisfy
the orthonormality and completeness conditions~(\ref{o2}) and
(\ref{o3}) and that (\ref{H=}) is also satisfied by (\ref{2-H}) --
(\ref{phi=}). Now consider the case that the tow-level Hamiltonian
(\ref{2-H}) has a real determinant. Then, Equation~(\ref{eg-va})
implies that either both eigenvalues are real or they are
complex-conjugates. Therefore, as a consequence of Theorem~1, we
have:
    \begin{itemize}
    \item[] {\bf Proposition:} Every nondegenerate (traceless) two-level Hamiltonian
    with real determinant is pseudo-Hermitian.
    \end{itemize}

Next, we construct the matrix representation of the operators
$L_n$ and $\eta_i$ and check that indeed the
factorization~(\ref{factor}) holds for the two-level
Hamiltonian~(\ref{2-H}) with real determinant. In order to do this
we consider the cases of real and complex eigenvalues separately.

\begin{itemize}
\item[] {\bf Case I.} $E$ is real (and positive)\\
In this case, according to (\ref{s1}) -- (\ref{alpha=}), $\alpha_1=\alpha_2=\sqrt{E}$ and
$\sigma^{(2)}=-\sigma^{(1)}=+1$. Therefore,
    \bea
    L&=&L(\alpha_1,\alpha_2)=\sqrt{E}(|\psi_1\kt\br\phi_1|+|\psi_2\kt\br\phi_2|)=
    \sqrt{E}~1,
    \label{2-L}\\
    \eta_1&=&-|\phi_1\kt\br\phi_1|+|\phi_2\kt\br\phi_2|,
    \label{2-eta1}\\
    \eta_2&=&|\phi_1\kt\br\phi_1|+|\phi_2\kt\br\phi_2|,
    \label{2-eta2}
    \eea
These in turn imply $\eta_1^{-1}=-|\psi_1\kt\br\psi_1|+|\psi_2\kt\br\psi_2|$ and
    \be
    L^\sharp L=\sqrt{E}L^\sharp=E\eta_1^{-1}\eta_2=
    E(-|\psi_1\kt\br\phi_1|+|\psi_2\kt\br\phi_2|)=H.
    \label{conf1}
    \end{equation}
\item[] {\bf Case II.} $E$ is not real\\
In this case, there is no real eigenvalues and we set $\alpha_1=1$, $\alpha_2=E$ and find
    \bea
    L&=&L(\alpha_1,\alpha_2)=|\psi_1\kt\br\phi_1|+E|\psi_2\kt\br\phi_2|,
    \label{2i-L}\\
    \eta_1&=&\eta_2=|\phi_1\kt\br\phi_2|+|\phi_2\kt\br\phi_1|,
    \label{2i-eta12}\\
    \eta_1^{-1}&=&|\psi_1\kt\br\psi_2|+|\psi_2\kt\br\psi_1|,
    \label{2i-eta1inv}\\
    L^\sharp&=&\eta_1^{-1}L\eta^2=-E|\psi_1\kt\br\phi_1|+|\psi_2\kt\br\phi_2|,
    \label{2-L-sharp}\\
    L^\sharp L&=&-E |\psi_1\kt\br\phi_1|+E|\psi_2\kt\br\phi_2|=H.
    \label{conf2}
    \eea
\end{itemize}
Equations~(\ref{conf1}) and (\ref{conf2}) confirm the validity of our results for arbitrary two-level Hamiltonians.

The matrix forms of the operators $\eta_1$, $\eta_2$, $L$ and $L^\sharp$ can be easily obtained by substituting (\ref{psi=}) and (\ref{phi=}) in (\ref{2-L}) -- (\ref{2-L-sharp}) in each case. As an illustrative example, we consider the two-level system associated with the two-component form of the classical equation of motion for a simple harmonic oscillator, namely $\ddot x(t)+\omega^2(t)x(t)=0$
where $\omega=\omega(t)$ is the frequency and a dot means a time-derivative. As discussed in \cite{jmp-99}, this equation may be written in the form of the Schr\"odinger equation $i\dot\Psi(t)=H(t)\Psi(t)$ for
    \be
    \Psi(t)=\left(\begin{array}{c}
    x(t)\\
    \dot x(t)\end{array}\right),~~~~~~~
    H_{\rm o}(t)=\left(\begin{array}{cc}
    0&i\\
    -i\omega(t)^2&0\end{array}\right).
    \label{osc}
    \end{equation}
Comparing (\ref{osc}) with (\ref{2-H}), we see that for the oscillator $a=0$, $b=i$, $c=-i\omega^2$,
$E=\omega$, and $N=2\omega^2$. As $\omega$ is real, (\ref{osc}) is an example of Case~1. For
this system, we have
    \bea
    |\psi_1\kt&=&\left(\begin{array}{c}
    -i\\
    \omega\end{array}\right),~~~~~~~
    |\psi_2\kt=\left(\begin{array}{c}
    \omega\\
    -i\omega^2\end{array}\right),
    \label{opsi=}\\ &&\nn\\
    |\phi_1\kt&=&\frac{1}{2}\,
    \left(\begin{array}{c}
    -i\\
    \omega^{-1}\end{array}\right),~~~~~~~~
    |\phi_2\kt=\frac{1}{2}\,
    \left(\begin{array}{c}
    \omega^{-1}\\
    -i\omega^{-2}\end{array}\right),
    \label{ophi=}\\ &&\nn\\
    \eta_1&=&\frac{1}{4\omega^2}\left(\begin{array}{cc}
    \omega^2(1-\omega^2)&i\omega(1+\omega^2)\\
    -i\omega(1+\omega^2)&1-\omega^2
    \end{array}\right),
    \label{oeta1}\\ &&\nn\\
    \eta_1^{-1}&=&\frac{1}{4\omega^2}\left(\begin{array}{cc}
    -1+\omega^2 & i\omega(1+\omega^2)\\
    -i\omega(1+\omega^2) & \omega^2(-1+\omega^2)
    \end{array} \right),
    \label{invoeta1}\\ &&\nn\\
    \eta_2&=&\frac{1}{4\omega^2}\left(\begin{array}{cc}
    \omega^2(1+\omega^2)&i\omega(1-\omega^2)\\
    -i\omega(1-\omega^2)&1+\omega^2\end{array}\right),\nn\\ &&\nn\\
    L&=&\sqrt{\omega}\left(\begin{array}{cc}
    1&0\\
    0&1\end{array}\right),~~~~~~~~~
    L^\sharp=\frac{1}{\sqrt\omega}\left(\begin{array}{cc}
    0&i\\
    -i\omega(t)^2&0\end{array}\right).\nn
    \eea
Clearly $L^\sharp L=L L^\sharp=H_{\rm o}$.

As a second check of our general results, we shall next show that the (non-Hermitian) oscillator Hamiltonian (\ref{osc}) can be obtained by intertwining the (Hermitian) spin Hamiltonian:
    \be
    H_{\rm s}=\omega(t)\sigma_3=\left(\begin{array}{cc}
    \omega(t) & 0\\
    0&-\omega(t)\end{array}\right).
    \label{spin}
    \end{equation}
This is interesting as there are well-known similarities between the quantum mechanical Hamiltonian for the simple harmonic oscillator and the spin Hamiltonian, e.g., for both systems the quantum dynamics is determined by the classical dynamics, \cite{nova}.  As we shall see pseudo-supersymmetry reveals a direct relationship between the two-level system associated with the classical simple harmonic oscillator and the (quantum) spin-half system described by the Hamiltonian~(\ref{spin}).

Letting $H_1=H_{\rm o}$ and $H_2=H_{\rm s}$ and using the notation of the preceding sections
and Equations~(\ref{eg-va}) -- (\ref{phi=}), we see that $|\psi_{n_0}^{(1)}\kt$,
$|\phi_{n_0}^{(1)}\kt$, $\eta_1$, and $\eta_1^{-1}$ are respectively given by (\ref{opsi=}), (\ref{ophi=}), (\ref{oeta1}), and (\ref{invoeta1}). In view of the diagonal form of (\ref{spin}),
we may choose
    \be
    |\psi_{1}^{(2)}\kt=|\phi_{1}^{(2)}\kt=\left(\begin{array}{c}
    0\\
    1\end{array}\right),~~~~~~~~~
    |\psi_{2}^{(2)}\kt=|\phi_{2}^{(2)}\kt=\left(\begin{array}{c}
    1\\
    0\end{array}\right),
    \label{pp-spin}
    \end{equation}
in which case $\eta_2=\eta_2^{-1}$ is just the identity matrix. Having chosen the complete biorthonormal systems $\{|\psi_{n_0}^{(i)}\kt,|\phi_{n_0}^{(i)}\kt\}$ and obtained $\eta_i$
and $\eta_i^{-1}$, we next compute
    \bea
    L&=&L(\alpha_1=\sqrt E,\alpha_2=\sqrt E)=
    \sqrt{\omega}\left(|\psi_1^{(2)}\kt\br\phi_1^{(1)}|+
    |\psi_2^{(2)}\kt\br\phi_2^{(1)}|\right)\nn\\
    &=&\frac{\sqrt{\omega}}{2}
    \left(\begin{array}{cc}
    \omega^{-1}&i\omega^{-2}\\
    i&\omega^{-1}\end{array}\right),
    \label{oL}\\&&\nn\\
    L^\sharp&=&\eta_1^{-1}L^\dagger\eta_2=
    \sqrt{\omega}
    \left(\begin{array}{cc}
    \omega & i\\
    -i\omega^2 &-\omega\end{array}\right),
    \label{oLsharp}\\&&\nn\\
    L^\sharp L&=&\left(\begin{array}{cc}
    0&i\\
    -i\omega(t)^2&0\end{array}\right)=H_{\rm o}=H_1,\nn\\ &&\nn\\
    LL^\sharp&=&\left(\begin{array}{cc}
    \omega(t) & 0\\
    0&-\omega(t)\end{array}\right)=H_{\rm s}=H_2.\nn
    \eea
The last two relations establish the fact that indeed $H_{\rm o}$ and $H_{\rm s}$ are pseudo-superpartner Hamiltonians.

\section{Conclusion}
In this article we derived some of the basic properties of
pseudo-supersymmetric quantum mechanics and explored its
consequences. Under the assumption of the diagonalizability of the
Hamiltonians and discreteness of their spectra, we showed that
indeed every pair of isospectral pseudo-Hermitian Hamiltonians
with identical degeneracy structure, except possibly for the zero
eigenvalue, define a pseudo-supersymmetric system. As
pseudo-Hermiticity and the presence of antilinear symmetries are
equivalent conditions, our results apply to the Hamiltonians
possessing antilinear symmetries, the typical example being the
$PT$-symmetric Hamiltonians. In fact, all the attempts made in the
literature to use Darboux's intertwining method in constructing
non-Hermitian Hamiltonians with a real spectrum may be viewed as
manifestations of pseudo-supersymmetry. Another area where
pseudo-supersymmetry is readily applied is in the study of
nondegenerate two-level Hamiltonians. Because all the
diagonalizable traceless matrix Hamiltonians with real determinant
are pseudo-Hermitian, these provide a class of quantum systems to
which our results apply generally. In particular, we checked the
$H=L^\sharp L$ factorization of all such Hamiltonians by explicit
calculations and showed how the two-level Hamiltonians describing
the classical dynamics of a simple harmonic oscillator and a
spin-half particle in a magnetic field are linked via a
pseudo-supersymmetry transformation.

In conclusion, we wish to point out that as far as the general
results of this article are concerned, lifting the discreteness
condition that we imposed on the spectrum of the Hamiltonian seems
not to lead to any major difficulties. In view of the analogy with
supersymmetry, we expect the presence of a continuous spectrum not
to diminish the utility of pseudo-supersymmetry. The similarity
between the algebraic structure of supersymmetry and
pseudo-supersymmetry suggests various generalizations of the
latter. For example, one may examine para-, ortho-, and fractional
pseudo-supersymmetry whose algebras are respectively obtained by
replacing the adjoint of the operators by their pseudo-adjoint in
the algebras of parasupersymmetry, orthosupersymmetry, and
fractional supersymmetry, \cite{susy2,ortho}. More generally, it
would be interesting to generalize the concept of a topological
symmetry \cite{npb-01} to quantum systems with a pseudo-Hermitian
Hamiltonian.

\section*{Acknowledgment}
This project was supported by the Young Researcher Award Program
(GEBIP) of the Turkish Academy of Sciences.

\ed